\documentclass[12pt]{article}
\usepackage{epsfig}
\usepackage{color}
\usepackage{amssymb,amsmath}
\usepackage [autostyle, english = american]{csquotes}
\MakeOuterQuote{"}

\newcommand{\bea}{\begin{eqnarray}}
\newcommand{\eea}{\end{eqnarray}}
\newcommand{\be}{\begin{equation}}
\newcommand{\ee}{\end{equation}}

 \makeatletter
\def\alt{\mathrel{\mathpalette\gl@align<}}
\def\agt{\mathrel{\mathpalette\gl@align>}}
\def\gl@align#1#2{\lower.6ex\vbox{\baselineskip\z@skip\lineskip\z@
\ialign{$\m@th#1\hfil##\hfil$\crcr#2\crcr\sim\crcr}}} \makeatother
\newcommand{\bigO}{\mathcal{O}}
\begin{document}
\begin{flushright}
%{\tt hep-ph/11mmxxx}\\
\end{flushright}
\vspace*{1.0cm}

\begin{center}
\baselineskip 20pt 
{\Large\bf 
Supersymmetric minimal $U(1)_X$ model at the TeV scale \\ 
with right-handed Majorana neutrino dark matter 
}
\vspace{1cm}

{\large 
Nobuchika Okada$^1$ and Desmond Villalba$^2$
}

\vspace{.5cm}

{\baselineskip 20pt \it
$^1$Department of Physics and Astronomy, \\
The University of Alabama, Tuscaloosa, AL 35487, USA\\
} 
{\baselineskip 20pt \it
$^2$Department of Chemistry and Physics, \\
University of Mary Washington, Fredericksburg , VA 22401, USA\\
}
\vspace{.5cm}

\vspace{1.5cm} {\bf Abstract}\\
\end{center}

We propose a supersymmetric extension of the minimal $U(1)_X$ model, along with a new $Z_2$-parity. 
One of the salient features of this model relates to how both the $U(1)_X$ gauge symmetry and R-parity are broken radiatively at the TeV scale by the VEV of a $Z_2$-even right handed sneutrino.
By assigning one right-handed neutrino $Z_2$-odd parity, it can remain a viable dark matter (DM) candidate, despite R-parity being broken. 
Furthermore, the DM relic abundance receives an enhanced annihilation cross section 
 due to the $U(1)_X$ gauge boson ($Z'$) resonance and is in agreement with the current observations. 
We have also found a complementarity that exists between the observed DM relic abundance and search results for the $Z^\prime$ boson resonance at the Large Hadron Collider (LHC), which further constrains the parameter space of our $U(1)_X$ model. 
Lastly, we consider a $SU(5)\times U(1)_x$ GUT extension and investigate the complementarities mentioned previously.

%>U(1)X extension and Z2 parity, particlesX
%>Natural way to break U(1)XX
%>Maintain DM candidate
%>Achieve acceptable DM abundance due to enhanced Z' resonanceX
%>Complimentarity between DM abundance and gauge coupling runningX
%>SU(5) easy to employ, with negligent effects due to mixing with U(1)X after SU(5) breaking.

\thispagestyle{empty}

%\bigskip
\newpage

\addtocounter{page}{-1}
\setcounter{footnote}{0}
%%%%%%%%%%%%%%%%%%%%%%%%%%
%\baselineskip 36pt
% Main body
%%%%%%%%%%%%%%%%%%%%%%%%%%
\baselineskip 18pt
%%%%%%%%%%%%%%%%%%%%%%%%%%
\section{Introduction}

The minimal supersymmetric (SUSY) extension of the Standard Model (MSSM) 
 is one of the prime candidates for physics beyond the Standard Model (BSM), 
 which naturally solves several problems in the SM, 
 in particular, the gauge hierarchy problem. 
In addition, a candidate for the cold dark matter, 
 which is missing in the SM, is also naturally 
 incorporated in the MSSM. 
The search for SUSY is one of the major directives
 of the Large Hadron Collider (LHC), which is operating at unprecedented luminosities, and is collecting data very rapidly. 
%As the search continues at the LHC for BSM physics the allowed parameter space for superpartner  Discovery of physics beyond the Standard Model in the near future 
 %is highly likely, and anticipated. 

Remarkably the MSSM can solve the gauge hierarchy problem and the dark matter problem. It is able to achieve the feat by mere virtue of it being supersymmetric. 
However it is clear that the SUSY extension 
 is not enough to solve the aforementioned problems 
 in addition to explaining neutrino phenomena. 
Both the observed solar and atmospheric neutrino oscillations, as well as long and short baseline experiments have established non-zero neutrino masses and mixings between different neutrino flavors~\cite{NuData}. 
Unlike the quark sector, the scale of neutrino masses 
 is very small and the different flavors are largely mixed. To make the MSSM a more viable description of nature we have no choice but to extend it, so that it incorporates neutrino masses and flavor mixings. 
The well known seesaw extension~\cite{seesaw} has garnered much support
 since it not only accounts for the neutrino mass but also 
 explains the smallness of the mass in a more "natural" way. 
Depending on the seesaw scale (typically the scale 
 of right-handed neutrinos) being, for example, 
 from 1 TeV to $10^{14}$ GeV, the scale of the neutrino 
 Dirac mass varies from 1 MeV (the electron mass scale) 
 to 100 GeV (the top quark mass scale).

As the $B-L$ (baryon number minus lepton number) 
 is an anomaly-free global symmetry in the SM, it can be easily gauged. The minimal $B-L$ model is the simplest gauged $B-L$ extension 
 of the SM~\cite{B-L}, where three generations of right-handed neutrinos 	 
 and a Higgs field with two units of the $B-L$ charge are introduced. 
The presence of the three right-handed neutrinos 
 is essential for canceling the gauge and gravitational anomalies. The general extension of the $B-L$ to the $U(1)_X$ model has been carried out \cite{U1X}, where the particle contents are the same except for the $U(1)_X$ charge assignment \cite{U1X2}. The $U(1)_X$ charge for a field is defined as a linear combination of $B-L$ and the hypercharge, $Q_x=Yx_H+Q_{B-L}$, where $x_H$ is a real parameter. In the limit of $x_H \rightarrow 0$ the MSSM $U(1)_{B-L}$ model is attained. As in the $B-L$ case, it has been shown that the $U(1)_X$ model is free of anomalies~\cite{U1XS}.

%In this model, the mass of right-handed neutrinos arises 
 %associated from the $U(1)_X$ gauge symmetry breaking. 

While the energy scale of the $U(1)_X$ gauge symmetry breaking is subject to some  phenomenological constraints, the energy  breaking scale is weakly constrained.
 Interestingly, through considerations of dark matter and collider physics, we have found that only a  small window in the parameter  region of a few TeV is allowed for the $U(1)_X$ model to remain viable.
 %In addition to the collider and cosmological constraints, the mass of right-handed neutrinos arises from the $U(1)_X$ gauge symmetry breaking. 
 % As the LHC continues to look for BSM particles, 
%Although the scale of the $U(1)_X$ gauge symmetry breaking is arbitrary 
% as long as phenomenological constraints are satisfied, 
% it is interesting to consider it at the %TeV scale~\cite{B-LTeV}. 
%For example, it has been recently pointed out~\cite{IOO} 
 %that when the classical conformal invariance is imposed 
% on the minimal $B-L$ model, the symmetry breaking scale 
% appears to be the TeV scale naturally. 
If this is the case, we can expect that all new particles in the model, 
 the $U(1)_X$  gauge boson $Z^\prime$, the $U(1)_X$ Higgs boson 
 and the right-handed neutrinos appear at the TeV scale, 
 which can be discovered at the LHC~\cite{LHCBL}. 
%The minimal $U(1)_X$ model also has interesting cosmological prospects 
% such as dark matter physics~\cite{OS} and baryogenesis~\cite{OOI2}. 

In this paper we investigate supersymmetric extension 
 of the minimal $U(1)_X$ model. 
It has been previously shown~\cite{KM} that an analogous  mechanism to radiative electroweak symmetry breaking in the MSSM exists for the case where the $U(1)_{B-L}$ symmetry is radiatively broken by the interplay 
 between large Majorana Yukawa couplings of right-handed neutrinos 
 and the soft SUSY breaking masses.
 %, a mechanism that is analogous 
% to radiative electroweak symmetry breaking in the MSSM. 
Employing the same mechanism naturally places the $U(1)_X$ symmetry breaking scale 
 at the TeV scale.

Despite this remarkable feature of the SUSY minimal $U(1)_{B-L}$ and our $U(1)_{X}$ model, 
 a more thorough analysis~\cite{fateR} indicated 
 that most of the $U(1)_{B-L}$ symmetry breaking parameter space 
 is occupied by non-zero vacuum expectation values 
 (VEVs) from right-handed sneutrinos. 
Therefore, the most likely scenario in the SUSY minimal $U(1)_X$ model 
 with the radiative $U(1)_X$ symmetry breaking, is that
 R-parity is violated in the vacuum. 
This means that the lightest superpartner (LSP) neutralino, 
 which is the conventional dark matter candidate 
 in SUSY models, becomes unstable and no longer remains
 a viable dark matter candidate.
As discussed in \cite{LSPgravitino}, 
 even though R-parity is broken, 
 an unstable gravitino if it is the LSP 
 has a lifetime longer than the age of the universe  
 and can still be the dark matter candidate. 
%Although this is an interesting possibility, 
% a mechanism of SUSY breaking mediations 
 %providing us with the LSP gravitino is limited, 
 %and we do not consider the LSP gravitino in this paper. 

A cogent framework for dark matter has been discussed 
 previously in the context of the (non-SUSY) minimal $B-L$ model~\cite{OS} and a $B-L$ MSSM~\cite{ZO}, 
 where a new $Z_2$-parity was introduced and one right-handed 
 neutrino was assigned odd $Z_2$-parity 
 while the other fields were assigned even $Z_2$. 
Calculation of the relic abundance of the $Z_2$-odd right-handed 
 neutrino showed that it could account for the observed 
 relic abundance, and therefore the dark matter in our universe. 
We mention this to emphasize that we are not introducing 
 any new particles in the current model.

In this paper, we apply the same idea to the SUSY generalization 
 of the minimal $U(1)_X$ model with the radiative $U(1)_X$ symmetry breaking, 
 and investigate the resulting phenomenology. 
What we discovered is that the $U(1)_X$ gauge symmetry 
 and R-parity are both broken at the TeV scale 
 by the non-zero VEV of a $Z_2$-even right-handed sneutrino, 
 for suitable regions of parameter space. 
Even in the presence of R-parity violation, 
 the $Z_2$-parity is still exact and the stability 
 of the $Z_2$-odd right-handed neutrino is guaranteed. 
Therefore, the $Z_2$-odd right-handed neutrino appears 
 to be a natural, stable dark matter candidate. 
We calculated the relic abundance of the $Z_2$-odd right-handed 
 neutrino and found that the resultant relic abundance was 
 in agreement with observations.

This paper is organized as follows. 
In the next section, we define the SUSY minimal $U(1)_X$ model 
 with $Z_2$-parity and introduce superpotential 
 and soft SUSY breaking terms relevant for our discussion. 
In Sec.~3, we perform a numerical analysis of the renormalization 
 group equation (RGE) evolution of the soft SUSY breaking masses
 of the right-handed sneutrinos and $U(1)_X$ Higgs fields 
 and show that the $U(1)_X$ gauge symmetry is radiatively broken 
 at the TeV scale. 
It will be shown that one $Z_2$-even right-handed sneutrino 
 develops a VEV and hence R-parity is also radiatively broken. 
In Sec.~4, we calculate the relic abundance of 
 the right-handed neutrino and identify the parameter region 
 consistent with the observed dark matter relic abundance. 
We also discuss phenomenological constraints 
 of the model in Sec.~5. 
In Sec.~6, we extend the model to the $SU(5)\times U(1)_X$ gauge group, and discuss SM gauge unification.
The last section is devoted to conclusions and discussions.

%%%%%%%%%%%%%%%%%%%%%%%%%%%%%%%%%%%%%%%%%%%%%%%%%%%%%%%%%%%%%%%%%%
\section{Supersymmetric Minimal $U(1)_X$ Model with $Z_2$-parity}
%%%%%%%%%%%%%%%%%%%%%%%%%%%%%%%%%%%%%%%%%%%%%%%%%%%%%%%%%%%%%%%%%%

The minimal $U(1)_X$ extended SM is based on the gauge group 
 $SU(3)_c\times SU(2)_L \times U(1)_Y \times U(1)_X$ 
 with three right-handed neutrinos and one Higgs scalar field 
 with $U(1)_X$ charge $2$, which is a singlet 
 under the SM gauge group%\footnote{
%Recently, a global $B-L$ extended model 
% with a dark matter candidate has been proposed~\cite{globalB-L}, 
 %which has some interesting differences to the gauged model. 
%}
. 
The $U(1)_X$ charges are defined as a linear combination of  $B-L$ and the hypercharge, $Q_x=Yx_H+Q_{B-L}$, where $x_H$ is a real parameter.
As far as the motivation to introduce three generations 
 of right-handed neutrinos ($N_i^c$) is concerned, 
 the introduction of the three generations of right-handed 
 neutrinos is in no way ad-hoc; 
On the contrary, once we gauge $U(1)_X$, their introduction 
 is forced upon us by the requirement of the gauge and 
 gravitational anomaly cancellations. 
The SM singlet scalar works to break the $U(1)_X$ 
 gauge symmetry by its Vacuum Expectation Value (VEV) and at the same time, 
 generates Majorana masses for right-handed neutrinos
 which then participate in the seesaw mechanism.

It is easy to supersymmetrize this model 
 and the particle contents are listed in Table~1\footnote{
It is possible to construct a phenomenologically viable 
 SUSY $B-L$ model without $\Phi$ and $\bar{\Phi}$~\cite{MMB-L}. 
}. 
The gauge invariant superpotential relevant for our discussion is given by 
\bea 
 W_{BL} = \sum_{i=2}^3 \sum_{j=1}^3 y_D^{ij} N^c_i L_j H_u 
         + \sum_{k=1}^3 y_k \Phi N^c_k N^c_k 
         - \mu_\Phi \bar{\Phi} \Phi, 
\label{WBL} 
\eea
 where the first term is the neutrino Dirac Yukawa coupling, 
 the second term is the Majorana Yukawa coupling for 
 the right-handed neutrinos, 
 and a SUSY mass term for the SM singlet Higgs fields 
 is given in the third term. 
Without loss of generality, we have worked in the basis 
 where the Majorana Yukawa coupling matrix is real and diagonal. 
Note that Dirac Yukawa couplings between $N^c_1$ and $L_j$ 
 are forbidden by the $Z_2$-parity, so that the lightest 
 component field in $N^c_1$ is stable, 
 as long as the $Z_2$-parity is exact. 

%%%%%%%%%%%%%%%%%%%%%%%%%%%%%%%%%%%%%%%%%%%%%%%
\begin{table}[t]
\begin{center}
\begin{tabular}{c|ccc|c|c|c}
chiral superfield & $SU(3)_c$ & $SU(2)_L$ & $U(1)_Y$ 
 & $U(1)_{X}$ & R-parity & $Z_2$ \\
\hline
$ Q^i   $  & {\bf 3}     & {\bf 2} & $+1/6$ & $(1/6)x_H+1/3$ & $-$ & $+$ \\ 
$ U^c_i $  & {\bf 3}$^*$ & {\bf 1} & $-2/3$ & $(-2/3)x_H-1/3$ & $-$ & $+$ \\ 
$ D^c_i $  & {\bf 3}$^*$ & {\bf 1} & $+1/3$ & $(1/3)x_H-1/3$ & $-$ & $+$ \\ 
\hline
$ L_i   $    & {\bf 1} & {\bf 2}& $-1/2$ & $(-1/2)x_H-1$   & $-$ & $+$ \\ 
$ N^c_1   $  & {\bf 1} & {\bf 1}& $  0 $ & $+1$   & $-$ & $-$ \\ 
$ N^c_{2,3} $  & {\bf 1} & {\bf 1}& $  0 $ & $+1$ & $-$ & $+$ \\ 
$ E^c_i $  & {\bf 1} & {\bf 1}& $ +1 $ & $x_H+1$     & $-$ & $+$ \\ 
\hline 
$ H_u$     & {\bf 1} & {\bf 2}   & $+1/2$ &  $ (1/2)x_H$ & $+$ & $+$ \\ 
$ H_d$     & {\bf 1} & {\bf 2}   & $-1/2$ &  $ (-1/2)x_H$ & $+$ & $+$ \\  
$ \Phi$    & {\bf 1} & {\bf 1}   & $ 0$   &  $-2$ & $+$ & $+$ \\  
$ \bar{\Phi}$ & {\bf 1} & {\bf 1}& $ 0$   &  $+2$ & $+$ & $+$ \\  
\end{tabular}
\end{center}
\caption{
Particle contents:  
In addition to the MSSM particles, 
 three right-handed neutrino superfields ($N^c_{1,2,3}$) 
 and two Higgs superfields ($\bar{\Phi}$ and $\Phi$) 
 are introduced. 
The $Z_2$-parity for $N^c_1$ is assigned to be odd.
$i=1,2,3$ is the generation index. 
}
\label{table1}
\end{table}
%%%%%%%%%%%%%%%%%%%%%%%%%%%%%%%%%%%%%%%%%%%%%%%

As we will discuss in the next section, 
 the $U(1)_X$ gauge symmetry is radiatively broken at the TeV scale, 
 and the right-handed neutrinos obtain TeV-scale Majorana masses. 
The seesaw mechanism\footnote{
As we will see in the next section, R-parity is also radiatively broken. 
In this case, the right-handed neutrinos mix with the $B-L$ gaugino 
 and fermionic components of $\bar{\Phi}$ and $\Phi$,  
 and the seesaw formula is quite involved.} 
 sets the mass scale of light neutrinos at 
 $m_\nu = m_{D}^T M_R^{-1} m_{D}=\frac{v_u^2}{2} y_{D}^T M_R^{-1}y_{D} $, 
 where $v_u$ is the VEV of the up-type Higgs doublet in the MSSM, 
 $M_R$ is the two-by-two mass matrix of the right-handed neutrinos,
 and $y_D$ is the two-by-three Dirac Yukawa coupling matrix from Eq.~(\ref{WBL}).
It is natural to assume that the mass of the heaviest light neutrino is 
 $m_\nu \sim \sqrt{\Delta m_{23}^2} \sim 0.05$ eV 
 with $\Delta m_{23}^2 \simeq 2.43 \times 10^{-3}$ eV$^2$ 
 being the atmospheric neutrino oscillation data~\cite{NuData}. 
Thus, we estimate $y_D \sim 10^{-6}$, and point out that
 such a small neutrino Dirac Yukawa coupling is negligible 
 in the analysis of RGEs.

Next, we introduce soft SUSY breaking terms 
 for the fields in the $U(1)_X$ sector: 
\bea 
  {\cal L}_{\rm soft}&=& 
- \left( \frac{1}{2} M_{X} \lambda_{X} \lambda_{X} + h.c.  \right)
- \left( \sum_{k=1}^3 m_{\tilde{N}^c_k}^2 |\tilde{N^c_k}|^2 
+ m_\Phi^2 |\Phi|^2 + m_{\bar \Phi}^2 |\bar{\Phi}|^2
  \right) 
\nonumber \\ 
&+& \left( B_\Phi \bar{\Phi} \Phi 
 + \sum_{k=1}^3 A_k \Phi \tilde{N^c_k} \tilde{N^c_k} + h.c. \right).
\eea
Here we have omitted terms relevant to the neutrino Dirac Yukawa couplings 
 since they are very small, i.e.  $\bigO(10^{-6})$ or smaller. 
For simplicity, in this analysis we consider
 the same setup as the constrained MSSM 
 and assume the universal soft SUSY breaking parameters, 
 $m_{\tilde{N}^c_k}^2 = m_\Phi^2 = m_{\bar \Phi}^2 = m_0^2$ 
 and $A_k = A_0$, 
 at the grand unification scale\footnote{
However, we do not necessarily assume grand unification 
 behind our model.  
In fact, it is very non-trivial to unify the $Z_2$-odd 
 right-handed neutrino with $Z_2$-even fields. 
}, $M_U =2 \times 10^{16}$ GeV.

Before closing this section, we comment on the uniqueness 
 of the $Z_2$-parity assignment from the phenomenological 
 point of view. 
One may find the $Z_2$-parity assignment ad-hoc, 
 but we cannot assign an odd-parity for any MSSM particles 
 because the parity forbids the Dirac Yukawa couplings 
 which is necessary to reproduce the observed fermion masses 
 and quark flavor mixings. 
As we will see in the next section, the scalars 
 $\Phi$ and $\bar{\Phi}$ develop non-zero VEVs 
 to break the $U(1)_X$ gauge symmetry, and 
 these fields should be $Z_2$-parity even 
 in order to generate the Majorana masses for the right handed neutrinos. 
Hence, we can assign $Z_2$-odd parity only 
 for right-handed neutrinos. 
Considering the fact that we need at least two right-handed neutrinos 
 to reproduce the observed neutrino oscillation data, 
 two right-handed neutrinos should be parity even 
 and be involved in the seesaw mechanism. 
As a result, we have assigned $Z_2$-parity odd 
 for only one right-handed neutrino as in Table~1. 
This $Z_2$-parity can be considered as an enhanced global symmetry, which becomes manifest after taking the Dirac Yukawa coupling of $N^c_1$ to zero.

%%%%%%%%%%%%%%%%%%%%%%%%%%%%%%%%%%%%%%%%%%%%%%%%%%%%%%%%%%
\section{Radiative $U(1)_X$ Symmetry Breaking and R-parity} 
%%%%%%%%%%%%%%%%%%%%%%%%%%%%%%%%%%%%%%%%%%%%%%%%%%%%%%%%%%

In the non-SUSY minimal $U(1)_X$ model, the $U(1)_X$ symmetry breaking 
 scale is determined by parameters in the Higgs potential 
 which can in general be at any scale as long as 
 the experimental constraints are satisfied. 
The LEP experiment has set the lower bound 
 on the $B-L$ symmetry breaking scale as 
 $m_{Z'}/g_{BL} \geq 6-7$ TeV~\cite{vBL}. 
The most recent LHC results for $Z'$ boson search 
 with 139 fb$^{-1}$ \cite{ATLAS:2019} 
 excluded the $B-L$ $Z^\prime$ gauge boson mass 
 $m_{Z'} \lesssim 5.1$ TeV. 
We see that the LHC bound is more severe 
 than the LEP bound. 
The SUSY extension of the model, however, 
 offers a very interesting possibility for constraining 
 the $B-L$ (and therby $U(1)_X$) symmetry breaking scale, as pointed out in \cite{KM}.

It is well-known that the electroweak symmetry breaking 
 in the MSSM is triggered by radiative corrections 
 to the up-type Higgs doublet mass squared 
 via the large top Yukawa coupling~\cite{REWSB}. 
Directly analogous to this situation, the $U(1)_X$ symmetry breaking 
 occurs through radiative corrections 
 with a large Majorana Yukawa coupling.

\begin{figure}[t]
\begin{center}
{\includegraphics[scale=1.0]{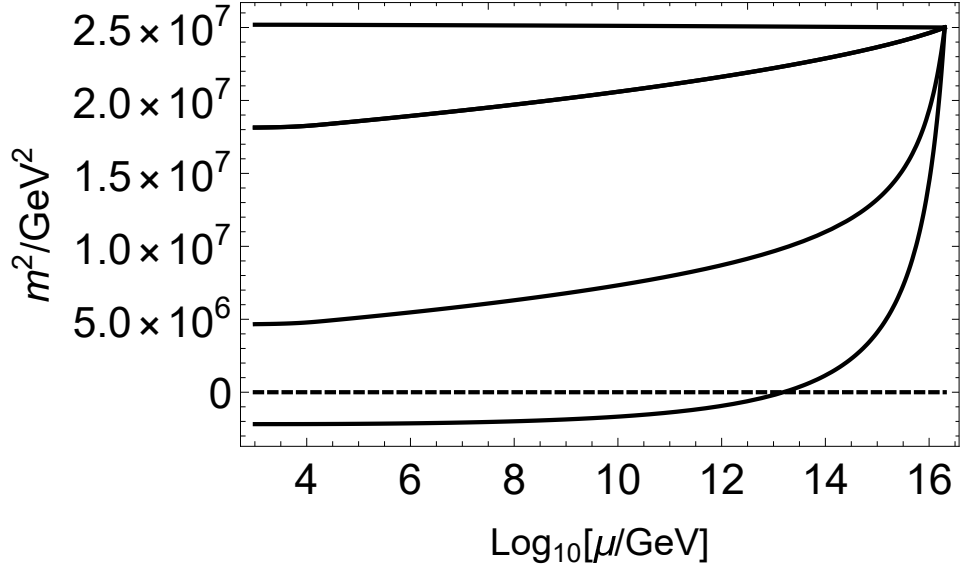}}
\caption{
The RGE running of the soft SUSY breaking masses, 
 $m_{\bar \Phi}^2$ (top curve), $m_{\tilde{N}^c_1}^2 =m_{\tilde{N}^c_2}^2$ (second from the top), 
 $ m_{\Phi}^2$ (third from the top) and $m_{\tilde{N}^c_3}^2$ (bottom curve) shown in Eq.~(\ref{RGE}).
}
\label{fig1}
\end{center}
\end{figure}

We consider the following RGEs for soft SUSY breaking terms 
 in the $U(1)_X$ sector~\cite{fateR, RGE2}:  
\bea 
 16 \pi^2 \mu \frac{d M_{X}}{d \mu} 
&=&  2(24 + 16 x_H + 11 x_H^2) g_{X}^2 M_{X}, \nonumber \nonumber \\ 
 16 \pi^2 \mu \frac{d m_{\tilde{N}^c_i}^2}{d \mu} 
&=&   8 y_i^2 m_\Phi^2 + 16 y_i^2 m_{\tilde{N}^c_i}^2
  + 8 A_i^2 - 8 g_{X}^2 M_{X}^2,  \nonumber \\
 16 \pi^2 \mu \frac{d m_{\Phi}^2}{d \mu} 
&=&   4 \left( \sum_{i=1}^3 y_i^2 \right) m_{\Phi}^2
 + 8 \sum_{i=1}^3 y_i^2 m_{\tilde{N}^c_i}^2 
 + 4 \sum_{i=1}^3 A_i^2 -32 g_{X}^2 M_{X}^2, \nonumber \\ 
 16 \pi^2 \mu \frac{d m_{\bar \Phi}^2}{d \mu} 
&=&  -32 g_{X}^2 M_{X}^2, \nonumber \\ 
 16 \pi^2 \mu \frac{d A_i}{d \mu}  
&=&  
\left(
  30 y_i^2 + 2 \sum_{j \neq i} y_j^2 - 12 g_{X}^2  \right) A_i
+ 4 y_i \left( \sum_{j \neq i} y_j A_j - 6 g_{X}^2 M_{X} 
 \right),  
 \label{RGE}
\eea  
 where RGEs for the gauge and Yukawa couplings are given by 
\bea 
  16 \pi^2 \mu \frac{d g_{X}}{d \mu} 
&=&  (24 + 16 x_H + 11 x_H^2) g_{X}^3 , \nonumber \\ 
  16 \pi^2 \mu \frac{d y_i}{d \mu}   
&=& y_i 
  \left( 
  10 y_i^2 + 2 \sum_{j \neq i} y_j^2 -12 g_{X}^2 
  \right). 
  \label{RGE2}
\eea
To illustrate the radiative $U(1)_X$ symmetry breaking, 
 we solve these equations from $M_U=2 \times 10^{16}$ GeV 
 to low energy, choosing $x_H=-0.8$ and the following boundary conditions. 
\bea 
&& g_{X}= 0.532, \; \; 
   y_1 = y_2 = 0.4, \; y_3= 2.5, \nonumber \\ 
&& M_{X}=1 \; {\rm TeV}, \; \; 
   m_{\tilde{N}^c_i} = m_{\Phi} = m_{\bar \Phi} = 5 \; {\rm TeV}, \; \; 
   A_i = 0.   
\label{BC}
\eea 
The RGE running of soft SUSY breaking masses 
 as a function of the renormalization scale 
 is shown in Fig.~\ref{fig1}. 
After the RGE running, $m_{\tilde{N}^c_3}^2$ becomes negative
 while the other squared masses remain positive. 
The negative mass squared of the right-handed sneutrino 
 triggers not only the $U(1)_X$ symmetry breaking 
 but also R-parity violation. 
Detailed analysis with random values of parameters 
 has shown~\cite{fateR} that 
 in most of the parameter region 
 realizing the radiative $B-L$ symmetry breaking, 
 R-parity is also  broken. 

%%%%%%%%%%%%%%%%%%%%%%%%%%%%%%%%%%%%%%%%%%%%%%%%
%%%%%%%%%%%%%%%%%%%%%%%%%%%%%%%%%%%%%%%%%%%%%%%%%

We now analyze the scalar potential with the soft SUSY breaking 
 parameters obtained from solving RGEs. 
Since the $U(1)_X$ symmetry breaking scale is set to be 26 TeV 
 in the following, we evaluate the RGE solutions at 26 TeV as follows: 
\bea 
&& g_{X}= 0.192, \; \; 
   y_1 = y_2 = 0.264, \; 
   y_3= 0.533, \nonumber \\ 
&& M_{X}=766 \; {\rm GeV}, \; \; 
   m_{\tilde{N}^c_1}^2 = m_{\tilde{N}^c_2}^2 
  = 1.83 \times 10^7 \; {\rm GeV}^2, \; \; 
   m_{\tilde{N}^c_3}^2 = -2.18 \times 10^6 \; {\rm GeV}^2, 
\nonumber \\ 
&& m_{\Phi}^2 = 4.91 \times 10^6 \; {\rm GeV}^2, \; \; 
   m_{\bar \Phi}^2 = 2.52 \times 10^7 \; {\rm GeV}^2, 
\nonumber \\ 
&& A_1 = A_2 = 30.4 \; {\rm GeV}, \; \; 
   A_3 = 36.5 \; {\rm GeV}. 
\label{input}
\eea 
The scalar potential for $\tilde{N}^c_3$, $\Phi$ and $\bar{\Phi}$ 
 consists of supersymmetric terms and soft SUSY breaking terms, 
\bea 
V= V_{SUSY} +V_{Soft},  
\eea  
where 
\bea 
&& V_{SUSY}= |2 y_3 \tilde{N}^c_3 \Phi|^2 + |\mu_\Phi \Phi|^2 
        + |y_3 (\tilde{N}^c_3)^2 - \mu_\Phi \bar{\Phi}|^2 
        + \frac{g_{X}^2}{2} 
\left(|\tilde{N}^c_3|^2 - 2 |\Phi|^2 + 2 |\bar{\Phi}|^2 \right)^2,  
\nonumber \\
&& V_{Soft} = 
  m_{\tilde{N}^c_3}^2 |\tilde{N}^c_3|^2 
 + m_{\Phi}^2 |\Phi|^2
 + m_{\bar \Phi}^2 |{\bar \Phi}|^2
 - \left( 
  A_3 \Phi \tilde{N}^c_3 \tilde{N}^c_3 + B_\Phi \bar{\Phi} \Phi + h.c. 
  \right). 
\eea
With appropriate values of $\mu_\Phi$ and $B_\Phi$, stationary conditions for the scalar potential can be found numerically.
% it is easy to numerically solve the stationary conditions 
% for the scalar potential. 
For example, we find (in units of TeV)  
\bea 
 \langle \tilde{N}^c_3 \rangle = \frac{12.5}{\sqrt 2}, \; \; 
 \langle \Phi \rangle =\frac{6.56}{\sqrt 2}, \; \;
 \langle \bar{\Phi} \rangle = \frac{9.29}{\sqrt 2} 
\eea
 for $\mu_\Phi=6.96$ TeV, $B_\Phi= 66.0$ TeV$^2$ 
 and the parameters given in Eq.~(\ref{input}). 
In this case, we have the $Z^\prime$ boson mass  
\bea 
  m_{Z'} = g_{X} v_{X} = 5 \; {\rm TeV}, 
\eea 
where 
\bea 
  v_{X	} = 
  \sqrt{
   2 \langle \tilde{N}^c_3 \rangle^2 
  + 8 \langle \Phi \rangle^2 + 8 \langle \bar{\Phi} \rangle^2} 
  = 26 \; {\rm TeV}
\eea 
 and the experimental lower bound 
 $v_{BL} \geq 6-7$ TeV~\cite{vBL} is satisfied.

In order to prove that the stationary point is actually 
 the potential minimum, we calculate the mass spectrum 
 of the scalars, $\tilde{N}^c_3$, $\Phi$ and $\bar{\Phi}$. 
By straightforward numerical calculations, 
 we find the eigenvalues of the mass matrix of the scalars 
 $\Re[\tilde{N}^c_3]$, $\Re[\Phi]$ and $\Re[\bar{\Phi}]$ as 
 $(13.7, 4.34, 4.75)$ in TeV, 
 while the mass eigenvalues for the pseudo-scalars 
 $\Im[\tilde{N}^c_3]$, $\Im[\Phi]$ and $\Im[\bar{\Phi}]$ 
 as $(0, 8.82, 12.8)$ in TeV. 
As expected, there is one massless eigenstate 
 corresponding to the would-be Nambu-Goldstone mode. 
The other right-handed sneutrino mass eigenvalues are given by 
\bea  
m_{\tilde{N}_{R i}}^2  = 
 m_{\tilde{N}^c_i}^2 
 + 4 y_i^2 \langle \Phi \rangle^2 
 - 2 y_i y_3 \langle \tilde{N}^c_3 \rangle^2 
 + 2 A_i \langle \Phi \rangle 
 + 2 y_i \mu_\Phi \langle \bar{\Phi} \rangle 
 + D_{X},  
\nonumber \\
m_{\tilde{N}_{I i}}^2  = 
 m_{\tilde{N}^c_i}^2 
 + 4 y_i^2 \langle \Phi \rangle^2 
 + 2 y_i y_3 \langle \tilde{N}^c_3 \rangle^2 
 - 2 A_i \langle \Phi \rangle 
 - 2 y_i \mu_\Phi \langle \bar{\Phi} \rangle 
 + D_{X}, 
\label{12} 
\eea
where $m_{\tilde{N}_{R i}}$ and $m_{\tilde{N}_{I i}}$ 
 ($i=1,2$) are the mass eigenvalues for scalars and pseudo-scalars, 
 respectively, and 
 $D_{X}=g_{X}^2 ( \langle \tilde{N}^c_3 \rangle^2 
-2 \langle \Phi \rangle^2  + 2 \langle \bar{\Phi} \rangle^2 ) $. 
We find $m_{\tilde{N}_{R 1}} = m_{\tilde{N}_{R 2}} = 5.58$ TeV 
 and $m_{\tilde{N}_{I 1}} = m_{\tilde{N}_{I 2}} = 5.16$ TeV. 
Since the fermion components in 
 $N^c_{2,3}$, $\Phi$ and $\bar{\Phi}$ and the $U(1)_X$ gauginos 
 are all mixed, it is quite involved to find the Majorana fermion 
 mass eigenvalues. 
Accordingly, the seesaw mechanism is realized in a very complicated way. 
Although we do not discuss the fermion spectrum in detail here, 
 our system with two right-handed neutrinos coupling to 
 the SM neutrinos provides many free parameters;
 enough to reproduce the observed neutrino oscillation data. 
On the other hand, the mass of the $Z_2$-odd 
 right-handed neutrino $N^c_{1}$ is simply given by\footnote{
It is generally possible to have a sneutrino DM candidate,
however, due to the LHC bound on the $Z^\prime$ gauge coupling,
the corresponding VEV must be large.
It would be difficult to tune the sneutrino mass to be half the $Z^\prime$ mass in order to achieve the correct relic abundance constraint discussed in section 4.} 
\bea  
 M_{N^c_1} = 2 y_1 \langle \Phi \rangle = 2.45 \; {\rm TeV} \simeq m_{Z^\prime}/2.
\eea

%%%%%%%%%%%%%%%%%%%%%%%%%%%%%%%%%%%%%%%%%%%%%%%%%%%%%%%%%%
\section{Right-handed Neutrino Dark Matter} 
%%%%%%%%%%%%%%%%%%%%%%%%%%%%%%%%%%%%%%%%%%%%%%%%%%%%%%%%%%
%%%%%%%%%%%%%%%%%%%%%%%%%%%%%%%%%%%%%%%%%%%%%%%%
\begin{figure}[ht]
\begin{center}
{\includegraphics[scale=0.8]{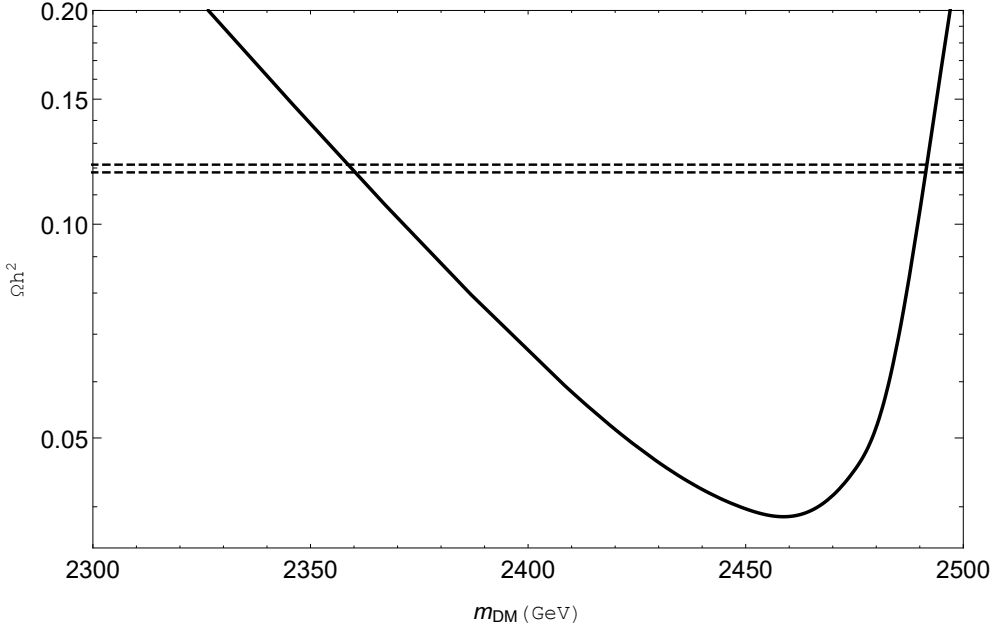}}
\caption{
The relic abundance of the dark matter right-handed neutrino 
 as a function of its mass for the $U(1)_X$ ($x_H=-0.8$) scenario. 
The dashed lines represent the upper and the lower bounds 
 on the dark matter relic abundance, $0.1183 \leq \Omega_{DM} h^2 \leq 0.1213$. 
}
\label{fig3}
\end{center}
\end{figure}
%%%%%%%%%%%%%%%%%%%%%%%%%%%%%%%%%%%%%%%%%%%%%%%%%
As we showed in the previous section, the $U(1)_X$ gauge symmetry 
 is radiatively broken at the TeV scale. 
Associated with this radiative breaking, 
 the right-handed sneutrino ${\tilde N}^c_3$ 
 develops  VEV and as a result, R-parity is also broken. 
Therefore, the neutralino is no longer the dark matter candidate. 
However, note that in our model the $Z_2$-parity is still exact, 
 by which the lightest $Z_2$-odd particle is stable and 
 can play the role of dark matter 
 even in the presence of R-parity violation. 
As is evident in the mass spectrum we found in the previous section, 
 the right-handed neutrino $N^c_1$ is the lightest $Z_2$-odd particle. 
In this section, we evaluate the relic abundance of 
 this right-handed neutrino dark matter candidate 
 and identify the parameter region(s) consistent 
 with the observations.

In \cite{OS}, the relic abundance of the right-handed neutrino 
 dark matter is analyzed in detail, 
 where annihilation processes through the SM Higgs boson 
 in the $s$-channel play the crucial role to reproduce 
 the observed dark matter relic abundance. 
In the non-SUSY minimal $B-L$ model, the right-handed neutrino 
 can have a sizable coupling with the SM Higgs boson 
 due to the mixing between the SM Higgs doublet and 
 the $B-L$ Higgs in the scalar potential. 
However, in supersymmetric extension of the $U(1)_X$ model 
 there is no mixing between the MSSM Higgs doublets 
 and the $U(1)_X$ Higgs superfields in the starting superpotential. 
Although such a mixing emerges through the neutrino Dirac Yukawa 
 coupling with the VEV of the right-handed sneutrino 
 ${\tilde N}^c_3$, it is very small because of the small 
 neutrino Dirac Yukawa coupling $y_D={\cal O}(10^{-6})$. 
Among several annihilation channels of 
 a pair  of the $Z_2$-odd right-handed neutrinos, 
 we find that the $s$-channel $Z^\prime$ boson exchange process gives 
 the dominant contribution.

Now we evaluate the relic abundance of the right-handed neutrino 
 by integrating the Boltzmann equation~\cite{KT}, 
\bea 
 \frac{dY_{N^c_1}}{dx}
 =-\frac{x \gamma_{Z^\prime}}{sH(M)} 
 \left[ \left(\frac{Y_{N^c_1}}{Y_{N^c_1}^{eq}}\right)^2-1 \right],  
\label{Boltmann}
\eea  
 where $Y_{N^c_1}$ is the yield (the ratio of the number density to 
 the entropy density $s$) of the $Z_2$-odd right-handed neutrino, 
 $Y_{N^c_1}^{eq}$ is the yield in thermal equilibrium, 
 temperature of the universe is normalized by the mass 
 of the right-handed neutrino $x=M/T$, 
 and $H(M)$ is the Hubble parameter at $T=M$. 
The space-time densities of the scatterings mediated 
 by the $s$-channel $Z^\prime$ boson exchange in thermal equilibrium 
are given by 
\bea 
\gamma_{Z^\prime} = 
 \frac{T}{64\pi^4} \int^\infty_{4 M^2} 
 ds \hat{\sigma}(s) \sqrt{s} K_1 \left(\frac{\sqrt{s}}{T}\right), 
\eea 
where $s$ is the squared center-of-mass energy, 
 $K_1$ are the modified Bessel function of the first kind, 
 and the total reduced cross section 
 for the process $N^c_1 N^c_1 \rightarrow Z' \rightarrow f \bar{f}$
 ($f$ denotes the SM fermions) is  
\bea 
 \hat{\sigma}_{Z^\prime}(s)
 = \frac{1}{24 \pi} g_{X}^4 
 \frac{\sqrt{s} \left( s - 4 M^2\right)^{\frac{3}{2}}}
 {\left(s-m_{Z'}^2 \right)^2 + m_{Z'}^2 \Gamma_{Z'}^2}F(x_H)%\times (13+16x_H+10x_H^2)
\eea 
 with the decay width of the Z$^\prime$ boson,  
\bea 
\Gamma_{Z'} = 
 \frac{g_{X}^2}{24 \pi} 
 \left[ F(x_H) + 2 \left( 1-\frac{4 M^2}{m_{Z'}^2} \right)^{\frac{3}{2}} 
 \theta \left( m_{Z'}^2/M^2 - 4 \right)  \right],
\label{width}
\eea 
where
\bea
F(x_H)=(13+16x_H+10x_H^2).
\eea
For simplicity, we have assumed that $y_1=y_2$ as in the previous section 
 and that the other particles (except for the SM particles) 
 are all heavy with mass $> m_{Z'}/2$. 
This assumption is consistent with the parameter choice 
 in our analysis below.

Now we solve the Boltzmann equation numerically. 
To solve the equation for the relevant domain, we inherit parameter values from those presented in the previous section which were already motivated as interesting values, 
\bea 
 g_{X}=0.192, \; \; m_{Z'}=5 \; {\rm TeV}, 
\eea
 while $M_{N^c_1}=M$ is taken to be a free parameter. 
With the asymptotic value of the yield $Y_{N^c_1}(\infty)$ 
 the dark matter relic density is written as 
\bea 
  \Omega h^2 =\frac{M s_0 Y_{N^c_1}(\infty)}{\rho_c/h^2}, 
\eea 
 where $s_0 = 2890$ cm$^{-3}$ is the entropy density 
 of the present universe, 
 and $\rho_c/h^2 =1.05 \times 10^{-5}$ GeV/cm$^3$ 
 is the critical density.
The result should be compared with the observations 
 at 2$\sigma$ level~\cite{WMAP7}
\bea 
 \Omega_{DM} h^2 = 0.1198 \pm 0.0015.  
\label{relicB}
\eea 
Fig.~\ref{fig3} shows the relic abundance of the right-handed 
 neutrino dark matter as a function of its mass. 
The dashed lines correspond to the upper and the lower bounds 
 on the dark matter relic abundance in Eq.~(\ref{relicB}). 
We find two solutions  
\bea 
 M \simeq 2359, \; \; 2492 \; {\rm GeV}.   
\eea 
It turns out from Fig.~\ref{fig3} that 
 in order to reproduce the observed relic abundance, 
 the enhancement of the annihilation cross section is necessary, 
 so that the mass of the dark matter should be close 
 to the $Z^{\prime}$ boson resonance point\footnote{
As the $Z^\prime$ boson partial decay width to a DM pair is negligibly small, the associated branching ratio is tiny (${\cal O}(0.1 \%$)).
At this stage, it is very challenging to understand the RHN DM existence directly through the Z’ boson measurement, but a future lepton collider such as the muon collider might be able to test our scenario with its TeV-scale collider energy and high precision measurements}. 
The dark matter mass $M=2492$ GeV coincides 
 with the value presented in the previous section. 
For a different parameter choice, the $Z_2$-odd right-handed sneutrino (the lighter of its scalar ($S$) or pseudo-scalar ($P$) components) can be the lightest $Z_2$-odd particle and a candidate for the dark matter, 
 instead of the right-handed neutrino. 
In this case, the main dark matter annihilation process is the co-annihilation process, $S P \to Z^\prime$.
Note that Eq.~(\ref{12}) indicates a sizable mass splitting between $S$ and $P$. 
This means that the co-annihilation process is not efficient even with the $Z^\prime$ resonance effect,
 since the number density of the particle that the dark matter particle co-annihilates with is suppressed
 much more than the dark matter number density. 

The RHN DM can scatter off with nuclei via $Z^\prime$ boson exchange. 
Since the RHN DM is a Majorana particle, only its interaction with nuclei is spin-dependent
  in the non-relativistic limit. 
We have estimated this spin-dependent cross section to be $\sigma_{SD} \sim 10^{-9}$ pb, 
which is far below the current upper bounds, $\sigma_{SD} \sim 10^{-5}$ pb for $m_{DM}={\cal O}$(1 TeV) 
\cite{LZ:2022lsv}.

%%%%%%%%%%%%%%%%%%%%%%%%%%
\section{LHC Constraints and Complementarity with Cosmological Bounds}
%%%%%%%%%%%%%%%%%%%%%%%%%%
The differential cross section for the process, $pp \to Z^\prime +X \to \ell^{+} \ell^{-} +X; \;\ell^{+} \ell^{-}=e^+ e^-/\mu^+ \mu^-$, 
   with respect to the dilepton invariant mass $M_{\ell \ell}$ is given by 
\begin{eqnarray}
 \frac{d \sigma}{d M_{\ell \ell}}
 =  \sum_{q, {\bar q}}
 \int^1_ \frac{M_{\ell \ell}^2}{E_{\rm LHC}^2} dx \, 
  \frac{2 M_{\ell \ell}}{x E_{\rm LHC}^2}  
 f_q(x, Q^2)  \,  f_{\bar q} \left( \frac{M_{\ell \ell}^2}{x E_{\rm LHC}^2}, Q^2
 \right) \times  {\hat \sigma} (q \bar{q} \to Z^\prime \to  \ell^+ \ell^-) ,
\label{CrossLHC}
\end{eqnarray}
where $Q$ is the factorization scale (we fix $Q= m_{Z^\prime}$, for simplicity),  
 $E_{\rm LHC}=13$ TeV is the center-of-mass energy of the LHC Run-2, 
 $f_q$ ($f_{\bar{q}}$) is the parton distribution function for quark (anti-quark), 
  and the cross section for the colliding partons is described as 
\bea 
{\hat \sigma}(q \bar{q} \to Z^\prime \to  \ell^+ \ell^-) =
\frac{\pi}{1296} \alpha_X^2 
\frac{M_{\ell \ell}^2}{(M_{\ell \ell}^2-m_{Z^\prime}^2)^2 + m_{Z^\prime}^2 \Gamma_{Z^\prime}^2} 
F_{q \ell}(x_H),  
\label{CrossLHC2}
\eea
where the function $F_{q \ell}(x_H)$ is given by 
\bea
   F_{u \ell}(x_H) &=&  (8 + 20 x_H + 17 x_H^2)  (8 + 12 x_H + 5 x_H^2),   \nonumber \\
   F_{d \ell}(x_H) &=&  (8 - 4 x_H + 5 x_H^2) (8 + 12 x_H + 5 x_H^2) 
\label{Fql}
\eea
for $q$ being the up-type ($u$) and down-type ($d$) quarks, respectively.
In our analysis, we employ CTEQ6L~\cite{CTEQ} for the parton distribution functions and 
  numerically evaluate the cross section of the dilepton production 
  through the $s$-channel $Z^\prime$ boson exchange. 
Since the right handed neutrino DM mass must be close to half of the $Z^\prime$ boson mass, 
  its contribution to the $Z^\prime$ boson decay width is negligibly small, 
  and thus the resultant cross section is controlled by only three free parameters, 
  $\alpha_X$, $m_{Z^\prime}$ and $x_H$.  
In interpreting the latest ATLAS results
\cite{ATLAS:2017} for the upper bound on 
   the cross section of the process $pp \to Z^\prime +X \to \ell^{+} \ell^{-} +X$, 
   we follow the strategy in Refs.~\cite{OO1, OO2, SO}: we first calculate the cross section of the process 
   by Eq.~(\ref{CrossLHC}) and then we scale our cross section result to find a $k$-factor ($k = 1.31$)  
   by which our cross section coincides with the SM prediction of the cross section presented in the ATLAS paper \cite{ATLAS:2017}. 
This $k$-factor is employed for all of our analysis.   
In this way, we find an upper bound on $\alpha_X$ as a function of $m_{Z^\prime}$ ($x_H$) for a fixed value of $x_H$ 
    ($m_{Z^\prime}$).  
  
The LEP experiments have searched for effective 4-Fermi interactions mediated by a $Z^\prime$ boson \cite{LEPdata}, 
and no significant deviation from the SM predictions have been observed. 
The LEP results are interpreted into a lower bound on $m_{Z^\prime}/\sqrt{\alpha_X}$ for a fixed $x_H$ value,  
  which means an upper bound on $\alpha_X$ as a function of $m_{Z^\prime}$ for a fixed $x_H$ value 
  similar to the constraints obtained from the LHC Run-2 results. 
For the minimal U(1)$_X$ model, the LEP bound on $m_{Z^\prime}/\sqrt{\alpha_X}$ has been obtained in Refs.~\cite{OO2,Das:2016zue}. 
Since the U(1)$_X$ charge assignment for the SM fermions in our model is the same as in the minimal model, 
  the LEP bound presented in Refs.~\cite{OO2,Das:2016zue} can be applied also to our model. 
Thus, we simply refer to the bound.    
We will see that the LHC constraints are much more severe than the LEP one for $m_{Z^\prime} \lesssim 5$ TeV.

To constrain the model parameter space further, we may also consider a theoretical upper bund on $\alpha_X$, 
   namely, the perturbativity bound on the gauge coupling.  
Recall that the beta function coefficient of the RGE
for the U(1)$_X$ gauge coupling from Eq.~(\ref{RGE2}) and the particle contents from \ref{table1} is given by
\bea 
    b_X=24 + 16 x_H + 11 x_H^2, 
\label{bX}    
\eea 
which is large compared to the SM $U(1)_Y$ RGE coefficient. 
To keep the running U(1)$_X$ gauge coupling $\alpha_X(\mu)$ in the perturbative regime up to the Planck scale ($M_{Pl}=1.22 \times 10^{19}$ GeV), 
    an upper bound on $\alpha_X$ at low energies can be derived. 
Solving the RG equation for the U(1)$_X$ gauge coupling at the one-loop level, 
  we find the relation between the gauge coupling at $m_{Z^\prime}$ 
  (denoted as $\alpha_X$ in our DM and LHC analysis) and the one at the Planck scale $\alpha_X(M_{Pl})$: 
\bea
  \alpha_{X}  =  \frac{\alpha_{X} (M_{Pl})}{1+ \alpha_{X} (M_{Pl}) \, \frac{ b_X }{2\pi} \ln \left[ \frac{M_{Pl}}{m_{Z^\prime}} \right]}.  
  \label{pert}
\eea 
For simplicity, we have set a common mass for all new particles to be $m_{Z^\prime}$. 
Effects of mass splittings are negligibly small unless the new particle mass spectrum is hierarchical. 
Imposing the perturbativity bound of  $\alpha_X(M_{Pl}) \leq 4 \pi$, we find an upper bound on $\alpha_X$ 
   for the fixed values of $m_{Z^\prime}$ and $x_H$. 

%%%%%%%%%%%%%%%%%%%%%%%%%%%%%%%%%%%%%%%%%%%%%%%%
\begin{figure}[t]
\begin{center}
{\includegraphics[scale=0.65]{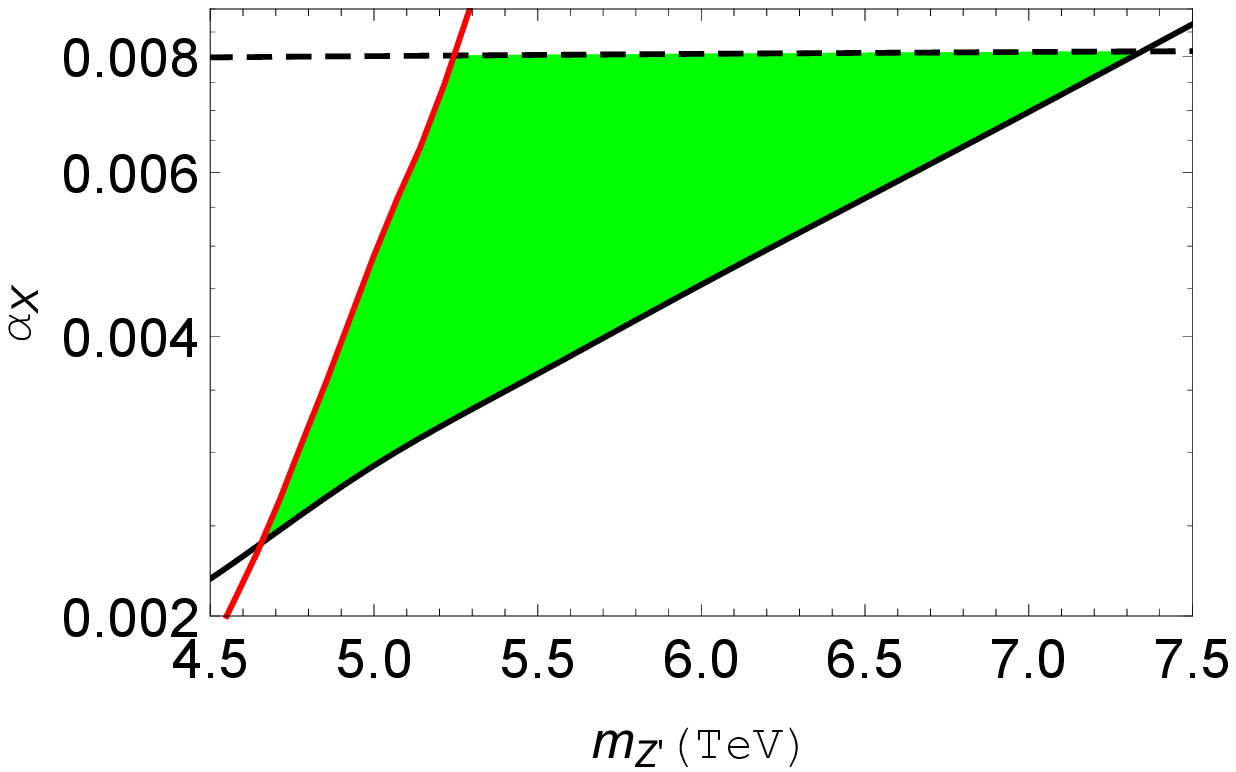}
}
\includegraphics[scale=0.65]{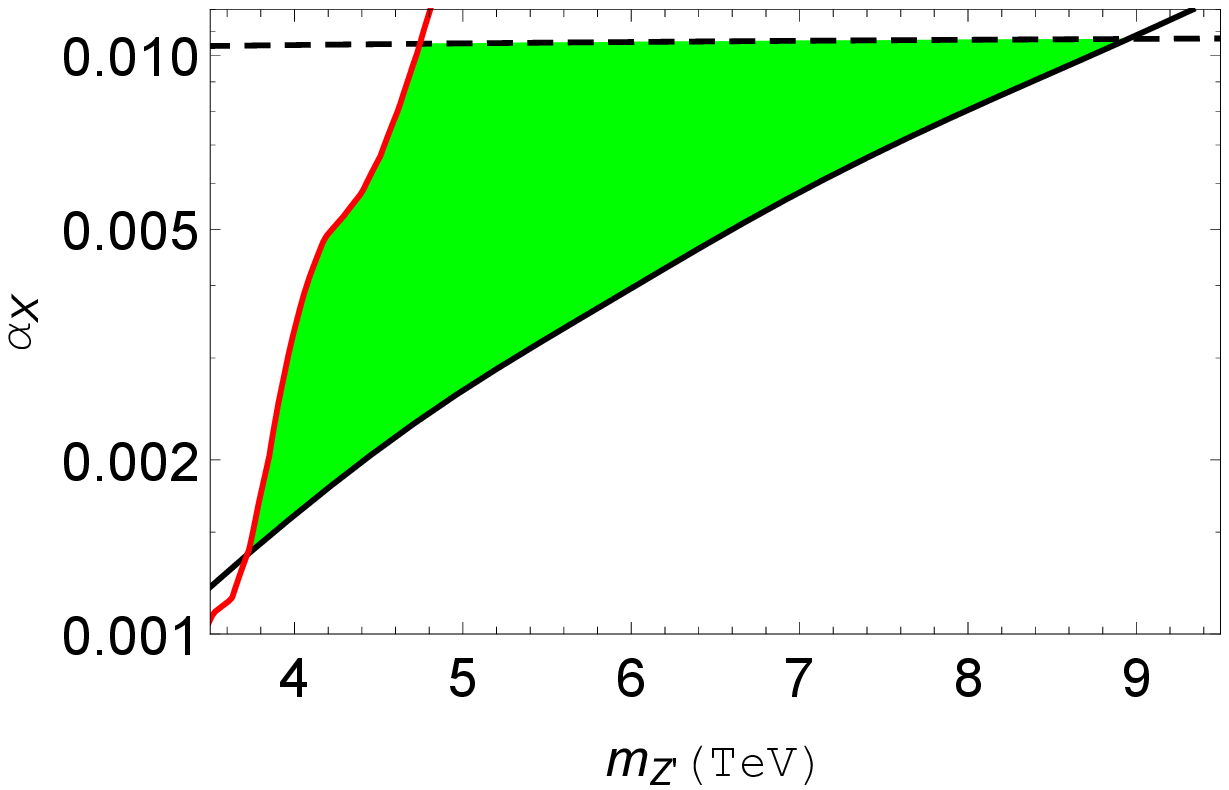}
\caption{
Combining the perturbativity constraints from solving the RGE's given by the dashed horizontal line (in black) and the DM relic abundance constraints previously discussed and shown by the lower solid curve (in black), yields a narrow allowed parameter region (shaded green) for the $U(1)_X$ model in the B-L ($SU(5)$) scenario, $x_H=0$ ($x_H=-0.8$) on the left (right). The diagonal solid line (in red) shows the updated LHC results. The LEP results are much less confining and well outside the confining region.
}
\label{fig:LHC}
\end{center}
\end{figure}
%%%%%%%%%%%%%%%%%%%%%%%%%%%%%%%%%%%%%%%%%%%%%%%%%

%%%%%%%%%%%%%%%%%%%%%%%%%%%%%%%%%%%%%%%%%%%%
%\begin{figure}[t]
%\begin{center}
%\includegraphics[width=0.49\textwidth, height=5.5cm]{XH1alphaVmZ}\;
%\includegraphics[width=0.49\textwidth, height=5.5cm]{XH2alphaVmZ}\\ 
%\;\;\;\;
%\includegraphics[width=0.49\textwidth, height=5.5cm]%{XH3alphaVmZ}
%\end{center}
%\caption{
%Allowed parameter regions for fixed $x_H$ values in ($%\alpha_X, m_{Z^\prime}$)-plane.  
%The solid lines are the cosmological lower bounds on $\alpha_X$ as a function of $m_{Z^\prime}$. 
%The (red) dashed and (red) dot-dashed lines are the upper bounds on $\alpha_X$ from the LHC and LEP results, respectively. 
%The perturbativity bounds on $\alpha_X$ are depicted by the dotted lines. 
%The (green) shaded regions satisfy all the constrains. 
%The top-left  (right) panel shows the result for a fixed $x_H = -1.2$ ($x_H =0$), and the bottom panel shows the result for $x_H = 2$.
%The LEP bound for $x_H = -1.2$ lies outside the range shown in the plot.  
%}
%\label{fig:LHC}
%\end{figure}
%%%%%%%%%%%%%%%%%%%%%%%%%%%%%%%%%%%%%%%%%%%% 

Let us now combine all constrains. 
We have obtained the lower bound on $\alpha_X$ from the observed DM relic abundance. 
On the other hand, the upper bound on $\alpha_X$ has been obtained 
  from the LHC results from the search for a narrow resonance, the LEP results 
  and the coupling perturbativity up to the Planck scale. 
Note that these constraints are complementary to narrow down the model parameter space.\footnote{
We see that  the LEP bound is always much weaker than the LHC bounds (for $m_{Z^\prime} \leq 5$ TeV)
  and the perturbativity bound. 
We have considered the LEP bound for completeness.   
} 
In Fig.~\ref{fig:LHC}, we show the combined results for $x_H=0$ ($x_H=-0.8$) in the left (right) panel. 
The (black) solid lines are the cosmological lower bounds on $\alpha_X$ as a function of $m_{Z^\prime}$. 
The (red) solid line is the upper bound on $\alpha_X$ from the LHC Run-2 results. 
The perturbativity bounds on $\alpha_X$ are depicted by the (black) dashed lines. 
The regions satisfying all the constraints are (green) shaded. 

\begin{figure}[t]
\begin{center}
{\includegraphics[scale=0.65]{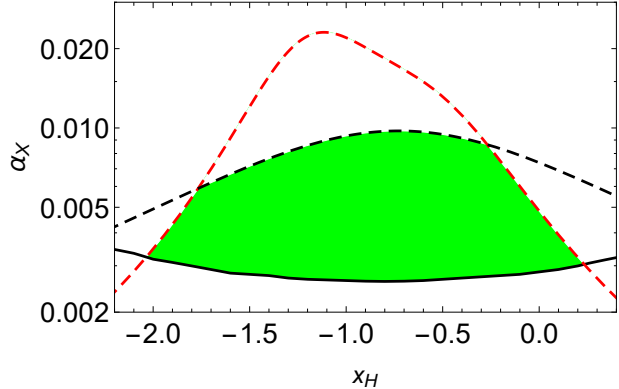}
}
\caption{
A scan over $x_H$ values for $m_{Z^\prime}=5 $ TeV combines the DM relic abundance constraints shown by the solid lower curve (in black), the perturbativity constraints shown by the second dashed curve from the top (in black), and the LHC data with 139 fb$^{-1}$ luminocity by the dashed upper curve (in red)~\cite{ATLAS:2019}. 
This narrow region between the constraints (green shaded) show the allowed values for $\alpha_X$. 
The value of $x_H=-0.8$ corresponds to the $SU(5)$ scenario, and is seen to be	 well within this allowed parameter region.
}
\label{fig:LHC2}
\end{center}
\end{figure}

Another interesting set of constraints on $\alpha_X$ to consider are found from combining a scan over $x_H$ values for the DM relic abundance bound, the purturbativity bound, and the latest LHC bounds. 
We show our combined results in Fig.~\ref{fig:LHC2} for $m_{Z^\prime}=5 $ TeV, 
  where the (red) dashed and (black) solid curves represent the LHC and DM relic abundance bounds, respectively,
and the (black) dashed curve illustrates the perturbativity bound on $\alpha_X$. 
The green-shaded region is allowed after combining all the bounds. 
The LHC bound shows the peak at $x_H \sim -1$. 
This is because the functions $F_{u \ell}$ and $F_{d \ell}$ in Eq.~(\ref{Fql}) have the minimum at $x_H \sim -1$. 
Similarly, the perturbative bound shows the peak at $x_H \sim -0.7$ since the beta function coefficient
 of Eq.~(\ref{bX}) has the minimum at $x_H \sim -0.7$. 
As expected, the LHC bound becomes weaker as we increase $m_{Z^\prime}$, leading to a wider green-shaded region.  
One can see from Fig.~\ref{fig:LHC2} that well within the allowed region (shaded green) sits the value of $x_H=-0.8$, which corresponds to the $SU(5)$ scenario discussed below. 
The fact that the $x_H=-0.8$ value lies in this region suggests that the $SU(5)$ scenario remains a viable description of nature, and as the LHC results are continually updated it will be interesting to see if the data continues to support the elegant $SU(5)$ case.

%%%%%%%%%%%%%%%%%%%%%%%%%%%%%%%%%%%%%%%%%%%%%%%%
%%%%%%%%%%%%%%%%%%%%%%%%%%%%%%%%%%%%%%%%%%%

\section{$SU(5)\times U(1)_X$ GUT Scenario}

Our setup can be readily extended to the $SU(5) \times U(1)_X$ gauge group. 
As has been previously shown in the non-SUSY $SU(5) \times U(1)_X$ setup in Ref.~\cite{U1X3}, this corresponds to the scenario where $x_H=-0.8$. 
Only with this choice for $x_H$ can the quarks and leptons be unified into the same supermultiplet, where the MSSM chiral superfields are arranged into the three generations of ${\bf 10}$ and ${\bf 5^*}$-representations under $SU(5)$. 
The $H_u$ and $H_d$ superfields are in the ${\bf 5}$ and ${\bf 5^*}$ representations, respectively. 
The $N_k^c$, $\Phi$, and $\bar{\Phi}$ superfields are all singlets under $SU(5)$.
An additional superfield that is neutral under $U(1)_X$ and in the ${\bf 24}$ representation of $SU(5)$ is required in order to break $SU(5)\times U(1)_X$ down to $SU(3)_C\times SU(2)_L\times U(1)_Y\times U(1)_X$. 
We consider the same $SU(5)$ breaking paradigm for the ${\bf 24}$ considered in \cite{SU5}, and we find that the unification scale for the SM gauge couplings occurs at $M_{GUT} \simeq 2 \times 10^{16}$ GeV. 
After $SU(5)$ has been broken down to the SM gauge groups at $M_{GUT}$, a kinetic mixing between the $U(1)_Y$ and  $U(1)_X$ gauge fields occurs due to the evolution of the RGEs. 

Following the procedure outlined in Ref.~\cite{U1X3}, the basis is chosen such that the gauge boson kinetic terms are all diagonalized. 
The covariant derivative of a field is given by

\bea
D_{\mu}=\partial_\mu - 
\begin{pmatrix}
Y & Q_X
\end{pmatrix}
\begin{pmatrix}
g_1 & g_{mix} \\
0 & g_X
\end{pmatrix}
\begin{pmatrix}
B_\mu \\
Z^\prime_\mu
\end{pmatrix},
\eea

where the Y and $Q_X$ are the $U(1)_Y$ and $U(1)_X$  field charges, respectively, 
$B_\mu$ and $Z^\prime_\mu$ are the SM $U(1)_Y$ and $U(1)_X$ gauge fields, 
and $g_1=\sqrt{5/3}~g_Y$ and $g_X$ are the $U(1)_Y$ and $U(1)_X$ gauge couplings. 
As a result of the original gauge kinetic mixing, a new parameter dubbed the "mixed gauge coupling" is introduced. 
In this chosen basis the RGE evolution of the SM $g_1$ gauge coupling remains unaffected, whereas the $g_X$ and $g_{mix}$ evolution evolve according to their coupled RGEs. 
At one-loop level, the RGEs for $\mu>\bigO ({\rm 	TeV})$ are given by 
\bea 
  16 \pi^2 \mu \frac{d g_{X}}{d \mu} 
&=&  g_{X}\left(
(24 + 16 x_H + 11 x_H^2) g_{X}^2 
+ 2(8+11 x_H)g_X g_{mix}
+ 11 g_{mix}^2
\right) , \nonumber \\ 
  16 \pi^2 \mu \frac{d g_{mix}}{d \mu}   
&=& g_{mix}\left(
(24 + 16 x_H + 11 x_H^2) g_{X}^2 
+ 2(8+11 x_H)g_X g_{mix}
+ 11 g_{mix}^2 \right) \nonumber \\ 
&+&\frac{6}{5}g_1^2\left(
(8+11 x_H) g_X+11 g_{mix}
\right). 
  \label{RGEX}
\eea

These RGEs encompass the effects of all particles in the theory present at the TeV scale. 
The RGEs in Eq.~(\ref{RGEX}) have been solved numerically with $g_{mix}=0$ and various values of $g_X$ at $\mu=M_{GUT}$. Regardless of the boundary value of $g_X$ at $M_{GUT}$, we have found that the ratio is always $g_{mix}/g_X \simeq 0.042$ at the TeV scale. 
The fact that this ratio is so small means that we can safely make the approximation to neglect the mixed gauge coupling in our analysis and set $g_{mix}=0$.
This approximation is consistent with all of our previous results attained for the $x_H=-0.8$ scenario.

\section{Conclusions and Discussions} 
%%%%%%%%%%%%%%%%%%%%%%%%%%%%%%%%%%%%%%%%%%%%%%%%%%%% 

The minimal gauged $U(1)_{X}$ model based on 
 the gauge group 
 $SU(3)_c \times SU(2)_L \times U(1)_Y \times U(1)_{X}$  
 is an elegant and simple extension of the Standard Model, 
 in which the right-handed neutrinos of three generations 
 are necessarily introduced for the gauge and gravitational 
 anomaly cancellations. 
The mass of right-handed neutrinos arises associated 
 with the $U(1)_X$ gauge symmetry breaking, 
 and the seesaw mechanism is naturally implemented. 
The supersymmetric extension of the minimal $U(1)_X$ model offers 
 not only a solution to the gauge hierarchy problem 
 but also a natural mechanism of breaking 
 the $U(1)_X$ symmetry at the TeV scale through
 the radiative $U(1)_X$ symmetry breaking. 
Although the radiative symmetry breaking at the TeV scale 
 is a remarkable feature of the model, R-parity is also 
 broken by non-zero VEV of a right-handed sneutrino. 
Therefore, the neutralino, which is the conventional 
 dark matter candidate in SUSY models, becomes unstable 
 and cannot play the role of the dark matter any more.

We have proposed the use of a $Z_2$-parity and 
 assigned an odd-parity to one right-handed neutrino. 
This parity ensures the stability of the right-handed neutrino 
 and hence the right-handed neutrino can remain a viable dark matter candidate 
 even in the presence of R-parity violation. 
In this way, no new particles need to be introduced as a candidate for dark matter. 
We have shown that for a parameter set, 
 the mass squared of a right-handed sneutrino 
 is driven to be negative by the RGE running. 
Analyzing the scalar potential with RGE solutions 
 of soft SUSY breaking parameters, we have identified 
 the vacuum where the $U(1)_X$ symmetry as well as R-parity 
 is broken at the TeV scale.

We have numerically integrated the Boltzmann equation for 
 the $Z_2$-odd right-handed neutrino and 
 found that its relic abundance is consistent with the observations. 
In reproducing the observed dark matter relic density, 
 an enhancement of the annihilation cross section 
 via the $Z^\prime$ boson $s$-channel resonance is necessary, 
 so that the dark matter mass should be close 
 to half of $Z^\prime$ boson mass.

Associated with the $U(1)_X$ symmetry breaking, 
 all new particles have TeV-scale masses, 
 which is being tested at the LHC in operation. 
Discovery of the $Z^\prime$ boson resonance at the LHC~\cite{LHCZ'}
 is the first step to confirm our model. 
Once the $Z^\prime$ boson mass is measured, the dark matter mass 
 is also determined in our model. 
If kinematically allowed, the $Z^\prime$ boson decays to 
 the dark matter particles with the branching ratio 
 $\sim 0.3$ \% (see Eq.~(\ref{width})). 
Precise measurement of the invisible decay width of $Z^\prime$ boson 
 can reveal the existence of the dark matter particle. 

We have also shown that the $SU(5)\times U(1)_X$ GUT scenario remains a possible description of nature by combining the constraints  on the $\alpha_X$ coupling from the perturbativity bound, LHC results on the process $pp \to Z^\prime +X \to \ell^{+} \ell^{-} +X; \;\ell^{+} \ell^{-}=e^+ e^-/\mu^+ \mu^-$, and DM relic abundance bound seen in Fig.~\ref{fig:LHC}. 
As seen in this figure, the lower mass bound for the $Z^\prime$ is around 5 TeV for this scenario. 
By scanning over $x_H$ values, one can see in Fig.~\ref{fig:LHC2} that the $x_H=-0.8$ value corresponding to $SU(5)$ remains in the narrow region of viability.

\section*{Acknowledgments}
%%%%%%%%%%%%%%%%%%%%%%%%%%%%%%%%%%
The work of N.O. is supported in part 
 by the DOE Grants, No. DE-SC0012447.

%%%%%%%%%%%%%%%%%%%%%%%%%%%%%%%%%%

%%%%%%%%%%%%%%%%%%%%%%%%%%%%%%%%%%
\end{document}